\documentclass[10pt,aps,prb,twocolumn,superscriptaddress,floatfix,english,twocolumn,amsmath,amssymmb,longbibliography]{revtex4-2}
\usepackage{amsfonts}
\usepackage{babel}
\usepackage{bm}
\usepackage{color}
\usepackage{comment}
\usepackage{dcolumn}
\usepackage{enumerate}
\usepackage{epsf}
\usepackage{subfigure}
\usepackage{epsfig,psfrag}
\usepackage{esint}
\usepackage{float}
\usepackage[T1]{fontenc}
\usepackage{graphicx}
\usepackage[normalem]{ulem}
\usepackage{latexsym}
\usepackage{relsize}
\usepackage[final]{hyperref} % adds hyper links inside the generated pdf file
\hypersetup{
	colorlinks=true,       % false: boxed links; true: colored links
	linkcolor=blue,        % color of internal links
	citecolor=blue,        % color of links to bibliography
	filecolor=magenta,     % color of file links
	urlcolor=blue         
}

\newcommand{\TBGpol}{\ensuremath{T_{\mathrm{BG}}^{(\mathrm{pol})}}}
\newcommand{\TBGtor}{\ensuremath{T_{\mathrm{BG}}^{(\mathrm{tor})}}}
\newcommand{\Srate}{\ensuremath{\Gamma}}
\newcommand{\f}{\ensuremath{f}}

\begin{document}

\title{Toroidal Fermi-surface geometry and phonon-limited transport in nodal-line semimetals}

\author{Aman Anand}
\email{Aman.Anand@citystgeorges.ac.uk}
\affiliation{Department of Mathematics, City St George's, University of London, London EC1V 0HB, United Kingdom}

\author{Alessandro De Martino}
\email{Alessandro.De-Martino.1@citystgeorges.ac.uk}
\affiliation{Department of Mathematics, City St George's, University of London, London EC1V 0HB, United Kingdom}	

\begin{abstract}
Nodal-line semimetals (NLSs) can display unconventional quasiparticle dynamics and 
charge transport properties due to their extended band degeneracy 
and the peculiar geometry of their Fermi surface. We consider electron–acoustic phonon scattering 
as the dominant relaxation mechanism and compute the 
quasiparticle decay rate and dc conductivity by solving the linearized semiclassical Boltzmann 
equation in a minimal model of a doped circular NLS. 
We find that the toroidal geometry of the Fermi surface gives rise to two parametrically distinct 
Bloch--Gr\"uneisen temperatures, associated with momentum transfers 
along the poloidal and toroidal directions, respectively. 
As a result, an intermediate temperature window opens between these two scales, 
in which the decay rate follows $\Gamma\propto T^2$, while the conductivity follows 
$\sigma\propto T^{-2}$. We also obtain the low- and high-temperature asymptotic behaviors, 
and discuss implications for ARPES and transport measurements in candidate NLS materials.
\end{abstract}

\maketitle

%%%%%%%%%%%%%

\section{Introduction}
\label{sec1}

Topological semimetals have emerged in recent years as a new platform for 
studying materials with symmetry- and topology-protected band crossings and the
unconventional transport properties that can result
~\cite{VafekAnnrev2014, Burkov2016, Armitage2018, Lv2021, SchoopCMreview2018}. 
Among them, nodal-line semimetals (NLSs)~\cite{Burkov2011, Fang2015, Chan2016, SYYang2018, Kwan2020}, 
in which valence and conduction bands touch along one-dimensional manifolds in momentum space, provide a particularly rich setting. 
Prototypical NLSs such as $\mathrm{ZrSiS}$~\cite{Schoop2016,Neupane2016,BBFU2019} host Dirac-like 
nodal lines and exhibit unusual phenomena,
including large magnetoresistance~\cite{XWang2016,Ali2016,Singha2017,Sankar2017}, 
high carrier mobilities~\cite{Sankar2017,Matusiak2017}, pronounced Zeeman splitting~\cite{Hu2017}, 
and enhanced quasiparticle masses near the nodal lines~\cite{Pezzini2018}, 
revealed by quantum oscillation experiments. 
Transport in these systems has been explored in various regimes,
showing unique signatures both in the quantum diffusive regime and 
in the presence of magnetic fields~\cite{MXYang2022, 
Skinner2017, Hu2019, Chen2019, Zhang2018, Pronin2021, HYang2018, Li2018,Pradhan2024, Poulomi2024, Rather2026}.

Impurity-limited transport has been analyzed in Ref.~\cite{Skinner2017}, 
where the conductivity was studied as a function of temperature, chemical potential, 
and impurity concentration. They noted that impurity potentials are strongly screened at elevated chemical potential. 
Experimentally, exceptionally clean samples of NLSs have been realized, like $\mathrm{ZrSiS}$ 
crystals displaying a large residual resistivity ratio, 
$\rho(300\,\mathrm{K})/\rho(2\,\mathrm{K}) = 288$~\cite{Singha2017}. 
These observations suggest the existence of a parameter window in temperature and chemical potential 
in which charge transport may be limited by inelastic scattering rather than disorder. 

In this regime, electron–phonon interactions \cite{RudenkoPRB2020, Fregoso2022, Lin2024, Suzumura2020} 
are expected to play a central role. In Ref.~\cite{RudenkoPRB2020}, these interactions were 
investigated with particular focus on optical phonons. 
The phonon spectrum of $\mathrm{ZrSiS}$ presented therein revealed that below 
$10\,\mathrm{meV}(\approx116\mathrm{K})$, 
only acoustic phonons are available for scattering. Signatures of electron–phonon scattering are 
observable in transport measurements as well as in angle-resolved photoemission spectroscopy (ARPES). 
ARPES studies~\cite{Bian2016, Hosen2017} on materials such as $\mathrm{PbTaSe}_2$ and the $\mathrm{ZrSi}X$ $(X = \mathrm{ S, Se, Te})$ 
family have revealed nodal-line dispersions and provided insights into quasiparticle lifetimes.

In conventional metals with large Fermi surfaces, the electron–phonon scattering rate 
exhibits two characteristic temperature regimes set by the Debye temperature~\cite{ZimanElectronPhonon, Ashcroft}. 
In semimetals, owing to the small size of the Fermi surface, the analogous crossover 
is controlled by a Bloch--Gr\"uneisen (BG) temperature, set by the maximal 
phonon momentum allowed for scattering on the Fermi surface~\cite{Fuhrer2010}.
For Weyl semimetals, electron--acoustic phonon scattering (both in bulk and in Fermi-arc states) 
and the role of the BG scales have been discussed in Refs.~\cite{Buccheri2021, Pereira2019}, while related Bloch–Gr\"uneisen physics has been explored experimentally in Dirac semimetals~\cite{Galeski2024}. 
However, a systematic analysis analogous to this for NLSs, 
where the Fermi surface acquires a toroidal geometry, is still lacking.

This motivates our study of electron-acoustic phonon scattering in NLSs in the clean limit. 
We focus on temperatures below the scale set by optical phonons and well below the chemical potential, 
so that interband processes are negligible. 
Notably, in non-symmorphic NLSs, the chemical potential can be tuned not only 
by doping but also via anisotropic strain~\cite{Topp2016}, providing experimental access to the regime we consider. 
While here we study how phonons affect electronic lifetime and transport, 
we note that the reciprocal problem of electron-induced renormalization of phonon properties 
has been addressed in Refs.~\cite{Singha2018, Xue2019,Lin2024,Zhao2024}.

We solve the linearized Boltzmann transport equation (BTE) exactly in the thin-torus limit by exploiting the fact that the leading-order scattering kernel depends only on angular differences on the torus.
Our main finding is that the toroidal geometry gives rise 
to two parametrically different BG temperatures, related to the two distinct radii of the Fermi surface. 
The coexistence of these two independent scales leads to a qualitatively new intermediate regime 
in which the phase space for electron–phonon scattering is modified, yielding a 
quasiparticle decay rate $\Gamma \propto T^2$ and a conductivity $\sigma \propto T^{-2}$. 
In addition, we find the usual behaviours $\Gamma \propto T^3$ and $\sigma \propto T^{-5}$ 
at low temperatures and $\Gamma \propto T$ with conductivity $\sigma \propto T^{-1}$ 
at high temperatures. A summary of these scaling regimes is provided in Table~\ref{tab:Table1}. 
The intermediate $T^2$ dependence has a transparent physical origin: in this temperature window, 
phonon momenta exceed the minor Fermi-surface radius but remain smaller than the major radius, 
so the scattering phase space along the poloidal direction has saturated while the toroidal direction 
is still thermally restricted, resulting in a partial saturation that is unique to the toroidal geometry.
Since a $T^2$ resistivity is frequently taken as a hallmark of electron–electron scattering~\cite{Baber1937, Ashcroft}, 
our results show that, in the circular NLS model considered here, the same scaling can also arise from 
intraband electron–phonon scattering, without invoking electron-electron interactions. 
We further observe an enhancement of the conductivity anisotropy at elevated temperatures.

The remainder of the paper is structured as follows. In Sec.~\ref{sec2}, we introduce 
the electronic and phonon models, the deformation-potential coupling, and the BG temperature scales. 
In Sec.~\ref{sec3} we briefly review the formulation of the linearized BTE
and derive integral expressions for the decay rate and the transport lifetime. 
Sections~\ref{sec4} and~\ref{sec5} present the asymptotic temperature dependences, 
together with numerical evaluations that confirm the predicted power laws. 
Finally, in Sec.~\ref{sec6} we conclude with a discussion of experimental implications and possible extensions.
Technical details are collected in several Appendices.

%%%%%%%%%%%%%

\section{Model}
\label{sec2}

In this section, we introduce the ingredients entering our transport calculation: 
a minimal two-band model for a circular NLS, acoustic phonons within an elastic continuum description, 
and their coupling via the deformation potential. 
We focus on the highly doped, low-temperature regime with chemical potential $\mu \gg k_B T$, 
where transport is dominated by intraband scattering processes 
in the conduction band. Throughout, we set $\hbar=1$. 
We consider systems in which the nodal line is approximately flat in energy, as realized, e.g., 
in Ca$_3$P$_2$~\cite{Chan2016} and strained YN~\cite{Huang2018}. Energy dispersion 
along the nodal contour would introduce an additional competing scale; its effect is discussed in Sec.~\ref{sec6}.
The toroidal Fermi surface that emerges at finite doping naturally introduces two distinct BG scales, 
discussed in Sec.~\ref{sec2:level4}, which set the relevant temperature regimes for scattering 
and transport discussed in Secs.~\ref{sec4} and~\ref{sec5}.

%%%%%%%%%%%%%

\subsection{Electronic model}
\label{sec2:level1}
\begin{figure}[t]
    \centering
    \includegraphics[width=1\linewidth]{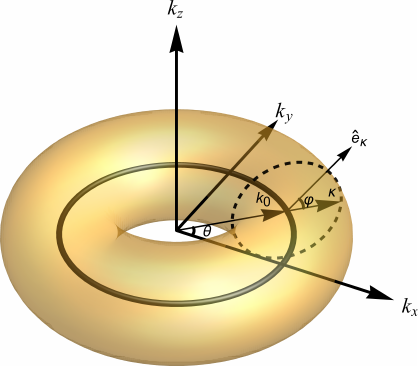}
    \caption{Shown are the toroidal Fermi surface, the associated toroidal coordinates 
    $(\kappa,\theta,\varphi)$, and the unit vector $\hat{e}_{\kappa}$, defined in Eq.~\eqref{Eq: toroidal-unit-vector-definition}.}
    \label{fig:torus1}
\end{figure}

We model low-energy quasiparticles in the vicinity of a circular nodal ring (NR)  
by the two-band effective Hamiltonian \cite{Burkov2011, Chan2016, MXYang2022}
\begin{equation}
    \mathcal H(\mathbf{k}) = vk_z \sigma_y + \lambda(k^2 - k_0 ^2) \sigma_z, 
    \label{Eq: OrigHamiltonian}
\end{equation}
where $k =\sqrt{k_x ^2 + k_y ^2 +k_z ^2}$, $\sigma_i$ are Pauli matrices, and $v,\lambda, k_0$ are material-dependent parameters. 
The nodal ring lies in the plane $k_z = 0$  at radius $k_0$. It is convenient to define
$v_0 = 2\lambda k_0$ and $\alpha = v/v_0$, where $v_0$ sets the characteristic in-plane velocity 
near the NR and $\alpha$ parametrizes the anisotropy along $k_z$ \cite{MXYang2022}. Representative values from DFT 
for Ca$_3$P$_2$
are $v \simeq 2.5$~eV\AA, $\lambda \simeq 4.34$~eV\AA$^2$,
and $k_0 = 0.206$ \AA$^{-1}$~\cite{Chan2016}, corresponding to $\alpha \simeq 1.4$ and $v_0 \simeq 2.6\times 10^5~\mathrm{m/s}$.
A small anisotropy may affect numerical prefactors and weakly rescale the BG temperatures, but we do not expect 
it to significantly affect the temperature exponents, therefore we set $\alpha = 1$ hereafter.

The electronic eigenstates are 
\begin{equation}
    |\Phi_{\mathbf{k},\zeta} ({\bf r})\rangle =  e^{i{\bf k}\cdot{\bf r}}|u_{\mathbf{k},\zeta} \rangle,
\end{equation}
where $|u_{\mathbf{k},\zeta} \rangle$ diagonalize $\mathcal H(\mathbf{k})$  with eigenvalues
\begin{equation}
    \varepsilon_{\mathbf{k},\zeta}  = \zeta \sqrt{v^2k_z ^2+ \lambda^2(k^2 - k_0 ^2)^2},
    \label{Eq: energy-eigenvals}
\end{equation}
and $\zeta = \pm $ labels conduction/valence bands. The corresponding Bloch spinors can be written as
\begin{equation}
    |u_{\mathbf{k},\zeta} \rangle = \mathcal{N}_{\mathbf{k},\zeta} \begin{pmatrix}
        ivk_z\\
         \lambda(k^2 -k_0 ^2) - \varepsilon_{\mathbf{k},\zeta } 
    \end{pmatrix} , 
    \label{BlochWavefunction}
\end{equation}
with normalization fixed by $\langle u_{\mathbf{k},\zeta'} |u_{\mathbf{k},\zeta} \rangle = \delta_{\zeta,\zeta'}$.
The group velocity $\mathbf{v}_\zeta(\mathbf{k}) = \partial_{\mathbf{k}}\varepsilon_{\mathbf{k},\zeta}$  is given by 
\begin{equation}
   \mathbf{v}_\zeta(\mathbf{k}) = \frac{2\lambda^2(\mathbf{k}^2-k_0^2)}{\varepsilon_{\mathbf{k},\zeta}}\mathbf{k} + 
  \frac{v^2}{\varepsilon_{\mathbf{k},\zeta}} k_z \hat{z}.
\end{equation}

For energies well below the Lifshitz scale $\varepsilon_{\mathbf{k},\zeta}\ll \lambda k_0^2$, 
constant-energy surfaces form a thin torus in momentum space (see Fig.~\ref{fig:torus1}). 
It is then convenient to introduce toroidal coordinates $(\kappa,\theta,\varphi)$ 
\begin{align}
     k_x & = (k_0 + \kappa \cos \varphi) \cos \theta, \label{Eq: torus-para-kx}\\
     k_y & = (k_0 + \kappa \cos \varphi) \sin \theta,\label{Eq: torus-para-ky}\\
     k_z & =  \kappa  \sin \varphi,\label{Eq: torus-para-kz}
\end{align}
where $\kappa$ is the minor radius, $\theta\in[0,2\pi)$ the toroidal angle (along the NR), 
and $\varphi\in[0,2\pi)$ the poloidal angle 
(around the tube cross-section). In the thin-torus regime $\kappa\ll k_0$, 
one may expand $k^2-k_0^2 \simeq 2k_0\kappa\cos\varphi$ and obtain the Dirac-like form
\begin{equation}
  \mathcal H(\kappa, \theta, \varphi) = v_0 \kappa (\cos \varphi \, \sigma_z + \sin \varphi \, \sigma_y),  
  \label{Eq: Hamiltonian-in-toroidal-coordinates}
\end{equation}
with dispersion 
\begin{equation}
    \varepsilon_{\mathbf{k},\zeta}=\zeta v_0\kappa.
    \label{lineardisp}
\end{equation}
Such linear behaviour  has been observed over a broad energy window in several NLS materials, 
including ZrSiS~\cite{Schoop2016, Neupane2016}.

In these coordinates, the spinors simplify to
\begin{equation}
  |u_{\mathbf{k},+} \rangle = \begin{pmatrix}
\cos \varphi /2\\
i\sin \varphi /2
\end{pmatrix},   \quad
  |u_{\mathbf{k},-} \rangle = \begin{pmatrix}
-\sin \varphi /2\\
i\cos \varphi /2
\end{pmatrix}.
\label{Eq: wavefunction-in-toroidal-coordinates}
\end{equation}
and the conduction-band group velocity becomes
\begin{align}
    \mathbf{v}_{+}(\mathbf{k})     & = v_0 \, \hat{e}_{\kappa}, 
\label{velocity}
\end{align}
where the unit vector $ \hat{e}_{\kappa}$ (shown in Fig.~\ref{fig:torus1}) reads
\begin{align}
    \hat{e}_{\kappa} =\cos \varphi \cos \theta \,\hat{x}+ \cos \varphi \sin \theta \,\hat{y}+ \sin \varphi \, \hat{z}.
    \label{Eq: toroidal-unit-vector-definition}
\end{align}
Since we focus here on temperatures $k_B T \ll \mu$, the valence band does not contribute, 
and we restrict attention to $\zeta=+$, suppressing the band index hereafter.

%%%%%%%%%%%%%

\subsection{Phonon model}
\label{sec2:level2}

We describe acoustic lattice vibrations within isotropic elastic continuum theory~\cite{Landau1986}. 
In this approximation, the phonon spectrum is determined by the longitudinal and transverse sound velocities $c_l$ and $c_t$. 
These velocities are typically much smaller than the electronic Fermi velocity; for example, in ZrSiS 
one finds $c_l \simeq 3.2\times 10^3~\mathrm{m/s}$~\cite{RudenkoPRB2020} 
and $v_0\simeq 6.5\times 10^5~\mathrm{m/s}$ \cite{Schoop2016}, so that $c_l/v_0\sim 10^{-2}$.

The displacement field ${\bf u}({\bf r},t)$ obeys the wave equation
\begin{align}
    \nabla_t ^2 {\bf u} = c_t ^2 \nabla ^2 {\bf u} + (c_l ^2 - c_t ^2)\nabla (\nabla \cdot {\bf u}), \label{Eq:phonon_eqn}
\end{align}
which, for harmonic time dependence ${\bf u}({\bf r},t) = {\bf u}({\bf r})e^{-i\Omega t}$, becomes
\begin{eqnarray}
    -\Omega^2{\bf u} = c_t ^2 \nabla ^2 {\bf u} + (c_l ^2 - c_t ^2)\nabla (\nabla \cdot {\bf u}). \label{Eq:phonon_main_eqn}
\end{eqnarray}
Its normal modes are one longitudinal and two transverse acoustic branches with dispersions $\Omega^{(l/t)}_{\mathbf{q}}=c_{l/t}|\mathbf{q}|$.

In the deformation-potential approximation, which we introduce in Sec.~\ref{sec2:level3}, 
only longitudinal lattice deformations contribute. So in what follows, 
we restrict ourselves to the longitudinal mode and we suppress the branch label 
in phonon operators and frequencies henceforth.
Upon quantization, the longitudinal displacement field can be expanded as
\begin{eqnarray}
    {\bf u}({\bf r}) = \int d^3\mathbf{q} \frac{e^{i{\bf q}\cdot{\bf r}}}{\sqrt{2\rho_M \Omega_{\bf q}}} 
  \hat{{\bf q}}  \left[ a_{\bf q}  - a_{-\bf q}^{\dagger} \right], 
    \label{Eq: quantized-displacement-field-solution}
\end{eqnarray}
where $\rho_M$ is the mass density of the medium, $a_{\bf q}^\dagger$, $a_{\bf q}$ 
are longitudinal phonon creation and annihilation operators, and $\hat{\mathbf q} = \mathbf q/|\mathbf q|$.

%%%%%%%%%%%%%

\subsection{Electron-phonon interaction}
\label{sec2:level3}

We consider electron–phonon interactions via the deformation potential,
\begin{equation}
    H_\mathrm{ep} = g_0 \nabla \cdot {\bf u}({\bf r}),
    \label{ephcoupling}
\end{equation}
where the coupling constant $g_0$ parametrizes the shift of the electronic band energy 
under local compression and is taken diagonal in spin--orbital space.
In principle, this coupling is renormalized by electronic screening, 
which we do not explicitly consider here. We therefore treat $g_0$ 
as an effective (screened) deformation-potential coupling, which may have a weak $\mu$ dependence. 

Since $\nabla\cdot\mathbf{u}=0$ for transverse modes, only longitudinal phonons contribute to the electron-phonon coupling. Other coupling mechanisms have also been explored, for example strain-induced effective gauge fields leading to a topological piezoelectric response in NLSs~\cite{Matsushita2020}. 
We neglect piezoelectric effects here because several NLSs of interest, 
including Ca$_3$P$_2$ and ZrSiS, are centrosymmetric and cannot exhibit conventional bulk piezoelectricity~\cite{Schoop2016, Chan2016, Kholkin2008}. 
In these systems, the deformation potential is expected to provide the dominant coupling between electron and acoustic phonons.

Substituting the mode expansion~\eqref{Eq: quantized-displacement-field-solution}, 
into $H_\mathrm{ep}$ and evaluating matrix elements between Bloch states $|\Phi_{\mathbf k}\rangle$ 
and $|\Phi_{\mathbf k'}\rangle$ yields
\begin{eqnarray}
    \langle \Phi_{\mathbf{k}'} | H_\mathrm{ep}| \Phi_{\mathbf{k}} \rangle =  
    {\mathcal{G}_{\mathbf{k},\mathbf{k}'}}\left( a_{\bf q}  - a_{-\bf q} ^{\dagger}  \right),
    \label{Eq: e-ph-matrix-element}
\end{eqnarray}
with momentum transfer ${\bf q} = {\bf k}' - {\bf k}$ and amplitude
\begin{equation}
{\mathcal{G}_{\mathbf{k},\mathbf{k}'}} = i \frac{g_0 \sqrt{\Omega_\mathbf{q}}}{c_l \sqrt{2\rho_M}} 
\langle u_{\mathbf{k}'} | u_{\mathbf{k}} \rangle . 
\label{Eq: StylishG}
\end{equation}
The pseudospin dependence enters through the Bloch spinor overlap $\langle u_{\mathbf{k}'}|u_{\mathbf{k}} \rangle$, 
which in toroidal coordinates follows from Eq.~\eqref{Eq: wavefunction-in-toroidal-coordinates} (for the conduction band)
as
\begin{equation}
    \langle u_{\mathbf{k}'}|u_{\mathbf{k}} \rangle = \cos \left(\frac{\varphi-\varphi'}{2} \right).
    \label{Eq: overlap}
\end{equation}

Assuming that phonons are in thermal equilibrium at temperature $T$, 
the transition rate from the electronic state $|\Phi_{\mathbf{k}}\rangle$  to 
$|\Phi_{\mathbf{k}'} \rangle$ follows from Fermi’s golden rule as
\begin{align}
W_{\mathbf{k}',\mathbf{k}} & = 2\pi |{\mathcal{G}_{\mathbf{k}',\mathbf{k}}}|^2 
\bigg{\{}  n_{\mathrm{B}}(\Omega_\mathbf{q} )  \, \delta (\varepsilon_{\mathbf{k}'} 
- \varepsilon_\mathbf{k} - \Omega_\mathbf{q} ) + \nonumber \\
&  + [n_{\mathrm{B}}(\Omega_\mathbf{q} ) + 1] \, \delta (\varepsilon_{\mathbf{k}'}  
- \varepsilon_\mathbf{k} + \Omega_\mathbf{q} ) \bigg{\}},  
\label{Eq:transition-rate-W} 
\end{align} 
where $n_{\mathrm{B}}(\Omega)=1/\left( e^{\beta \Omega} -1 \right)$ is the Bose-Einstein 
distribution function, with $\beta=1/k_BT$.
The two terms in Eq.~\eqref{Eq:transition-rate-W} describe phonon absorption and emission, 
respectively. These rates enter the collision integral of the Boltzmann equation in 
Sec.~\ref{sec3}.

%%%%%%%%%%%%%

\subsection{Temperature scales}
\label{sec2:level4}

In conventional metals with large Fermi surfaces, the crossover between low- 
and high-temperature regimes of electron-phonon scattering is usually controlled 
by the Debye temperature. In semimetals, instead, the crossover is controlled by the BG temperature, 
set by the maximal phonon momentum that can scatter carriers across the Fermi surface.
For a toroidal Fermi surface, there are two independent momentum scales, corresponding 
to the minor and major radii, and thus two BG scales. We define 
\begin{align}
    k_{\mathrm{B}} \TBGpol  &= 2c_l  \kappa_{\rm{F}},\\
    k_{\mathrm{B}} \TBGtor  &= c_l  k_0,
\end{align}
where $k_{\mathrm{B}}$ is the Boltzmann constant and $\kappa_\text{F}=\mu/v_0$ is the minor radius 
of the toroidal Fermi surface at chemical potential $\mu$. 
The scale $T_{\mathrm{BG}}^{(\mathrm{pol})}$ is associated with momentum transfer along the poloidal direction,
set by $\kappa_F$, while $T_{\mathrm{BG}}^{(\mathrm{tor})}$ reflects the toroidal extent along the NR, set by $k_0$. 

For illustration, in PbTaSe$_2$, a NLS with strong spin–orbit coupling~\cite{Bian2016},
one has $k_0 \simeq 0.2~\text{\AA}^{-1}$ and $c_l \simeq 3.2\times 10^3~\mathrm{m/s}$~\cite{MuhasinReza2023}, 
yielding
\begin{equation}
    \TBGtor 
    = \frac{ c_l k_0}{k_{\mathrm B}} \simeq 49~\mathrm{K}.
    \label{Eq: TBG-PbTaSe2}
\end{equation}
For ZrSiS, the nodal line is approximately diamond-shaped, but taking $k_0~\simeq ~0.3~\text{\AA}^{-1}$~\cite{Muller2020} 
and $c_l\simeq 6.8\times 10^3~\mathrm{m/s}$~\cite{RudenkoPRB2020} gives
\begin{equation}
    \TBGtor
    = \frac{ c_l k_0}{k_{\mathrm B}} \simeq 156~\mathrm{K}.
    \label{Eq: TBG-ZrSiS}
\end{equation}
The Debye temperatures in these materials are typically several times larger than
$T_{\mathrm{BG}}^{(\mathrm{tor})}$ \cite{Zhang2016, Sankar2017}, 
justifying the use of long-wavelength acoustic phonons with linear dispersion 
in the temperature ranges of interest.

In our model, we assume a thin torus, $\kappa_F\ll k_0$, which implies the hierarchy
\begin{equation}
  \TBGpol / \TBGtor \sim \kappa_\mathrm{F}/k_0 \ll 1.
\end{equation}
This naturally leads to three temperature regimes that will control 
both quasiparticle decay rate and transport scattering rate: 
(i) $T\ll  \TBGpol$, (ii) $\TBGpol \ll T\ll \TBGtor$, and (iii) $T\gg \TBGtor$. 

In the next section, we formulate the semiclassical BTE using the transition rates 
in Eq.~\eqref{Eq:transition-rate-W} and derive expressions for the decay rate and transport lifetime in 
these regimes.

%%%%%%%%%%%%%

\section{Boltzmann Equation}
\label{sec3}

In this section, we briefly review the formulation of semiclassical charge transport in the presence of 
electron-phonon scattering within the BTE framework~\cite{ZimanElectronPhonon}. 
We then solve the linearized BTE in terms of transport lifetimes $\tau^i_{\mathrm{tr}}$ ($i=x,y,z$), 
which we use to evaluate the conductivity tensor in Sec.~\ref{sec5}.
The same scattering kernel, without the transport weighting factor,  
also gives the single-particle decay rate $\Gamma$, discussed in Sec.~\ref{sec4}.

\subsection{Linearized BTE}
\label{sec:BTE_linearized}

In the presence of slowly varying external fields, the semiclassical BTE for the
electron distribution function 
$f(\mathbf{r},\mathbf{k},t)$ reads~\cite{Girvin2019ModernCM}
\begin{equation}
    \frac{\partial f}{\partial t}
    + \mathbf v(\mathbf k)\cdot\nabla_{\mathbf r} \mathbf{f}
    + \mathbf F_{\mathrm{ext}}\cdot \nabla_{\mathbf k} f
    =  \mathcal{I}_{\mathbf{k}}, 
    \label{Eq:BTE1}
\end{equation}
where $\mathcal{I}_{\mathbf{k}}$ is the collision integral, 
$\mathbf{v}(\mathbf{k})$ is the electron group velocity and $\mathbf{F}_{\mathrm{ext}} = e\mathbf{E}$ 
for a uniform electric field with $e$ the electron charge.
For a homogeneous steady state ($\partial_t f=\nabla_{\mathbf{r}}f=0$), driven by a weak uniform electric
field $\mathbf E$, Eq.~\eqref{Eq:BTE1} becomes
\begin{equation}
   e\mathbf{E}\cdot \nabla_{\mathbf{k}} f_{\mathbf{k}}
   =  \mathcal{I}_{\mathbf{k}}.
   \label{Eq:BTE2}
\end{equation}
The collision integral for scattering between states $\mathbf{k}$ and $\mathbf{k}'$ 
mediated by phonons can be written as
\begin{equation}
    \mathcal I_{\mathbf{k}}
    = \int \frac{d^3\mathbf{k}'}{(2\pi)^3}
      \bigl\{
        W_{\mathbf k,\mathbf k'} f_{\mathbf k'}(1-f_{\mathbf k})
        - W_{\mathbf k',\mathbf k} f_{\mathbf k}(1-f_{\mathbf k'})
      \bigr\},
    \label{Eq:BTE3}
\end{equation}
where $W_{\mathbf k',\mathbf k}$ is the transition rate from
$\mathbf k$ to $\mathbf k'$ obtained from Fermi's golden rule [cf. Eq.~\eqref{Eq:transition-rate-W}].

We now linearize the BTE around equilibrium by writing
\begin{equation}
    f_{\mathbf k}
    = n_{\mathrm F}(\varepsilon_{\mathbf k})
      - \varphi_{\mathbf k}\,
        \frac{\partial n_{\mathrm F}(\varepsilon_{\mathbf k})}
             {\partial\varepsilon_{\mathbf k}},
    \label{Eq: linearizing-distribution-function}
\end{equation}
where $n_{\mathrm F}(\varepsilon)=1/\left(e^{\beta(\varepsilon-\mu)}+1\right)$ 
is the Fermi-Dirac distribution function at chemical potential $\mu$,
and $\varphi_{\mathbf k}$ is a function parametrizing the deviation from equilibrium.
Standard manipulations using detailed balance (summarized in Appendix~\ref{App:Boltzmann}) recast the linearized collision integral as
\begin{equation}
\mathcal{I}_{\mathbf{k}}
=
\frac{\partial n_\text{F}(\varepsilon_{\mathbf{k}})}{\partial \varepsilon_{\mathbf{k}}}\,
\mathcal{J}_{\mathbf{k}},
\quad
\mathcal{J}_{\mathbf{k}}
\equiv
\int \frac{d^3\mathbf{k}'}{(2\pi)^3}\,
\mathcal{W}_{\mathbf{k}',\mathbf{k}}
\big( \varphi_{\mathbf{k}}-\varphi_{\mathbf{k}'}\big),
\label{Eq:BTE4}
\end{equation}
where the symmetric kernel $\mathcal W_{\mathbf k',\mathbf k}$ is given by
\begin{widetext}
\begin{equation}
    \mathcal W_{\mathbf k',\mathbf k}
    = 2\pi \bigl|\mathcal G_{\mathbf k',\mathbf k}\bigr|^2
      \bigl\{
        \bigl[n_{\mathrm B}(\Omega_{\mathbf q})
              + n_{\mathrm F}(\varepsilon_{\mathbf k} + \Omega_{\mathbf q})\bigr]
        \delta(\varepsilon_{\mathbf k'} - \varepsilon_{\mathbf k} - \Omega_{\mathbf q})
        +
        \bigl[n_{\mathrm B}(\Omega_{\mathbf q}) + 1
              - n_{\mathrm F}(\varepsilon_{\mathbf k} - \Omega_{\mathbf q})\bigr]
        \delta(\varepsilon_{\mathbf k'} - \varepsilon_{\mathbf k} + \Omega_{\mathbf q})
      \bigr\},
    \label{Eq: StylishW1}
\end{equation}
\end{widetext}
with $\mathbf q = \mathbf k' - \mathbf k$, and $\Omega_{\mathbf{q}}=c_l|\mathbf{q}|$. 
Inserting Eq.~\eqref{Eq:BTE4} into Eq.~\eqref{Eq:BTE2} yields the linearized BTE
\begin{equation}
   e \mathbf E\cdot \mathbf v(\mathbf k)
    = \mathcal J_{\mathbf k}.
    \label{Eq:BTE5}
\end{equation}

In the rest of this paper, we specialize to the degenerate, thin-torus regime. 
Specifically, we assume $k_BT\ll \mu$ so that transport is controlled by quasiparticles at the Fermi surface, 
and we set $-\partial_\varepsilon n_\mathrm{F}(\varepsilon) \simeq \delta(\varepsilon -\mu)$. Moreover, 
since $c_l\ll v_0$, typical phonon energies are small compared with electronic scales, 
so scattering is quasi-elastic ($\Omega_\mathbf{q}\ll \mu$ for thermal phonons), 
and we therefore omit the phonon energy in the energy-conserving $\delta$-functions 
in the scattering rates~\eqref{Eq: StylishW1}.
The combinations of Fermi and Bose functions evaluated at $\varepsilon_{\mathbf{k}} = \mu$, 
simplify to
\begin{align}
     n_{\mathrm{B}}(\Omega_\mathbf{q} ) & + n_{\mathrm{F}}( \varepsilon_{\mathbf{k}}+ \Omega_\mathbf{q} ) =
\frac{1}{\sinh (\beta\Omega_\mathbf{q} )}  \nonumber \\
     & = n_{\mathrm{B}}(\Omega_\mathbf{q} ) +1- n_{\mathrm{F}}(\varepsilon_{\mathbf{k}} - \Omega_\mathbf{q} ),
     \label{Eq: Fermi-Bose-function-simplification}
\end{align}
so that the kernel in Eq.~\eqref{Eq: StylishW1} takes the form  
\begin{equation}
\mathcal W_{\mathbf k',\mathbf k}= \widetilde{\mathcal W}_{\mathbf k',\mathbf k}  \, 
\delta( \varepsilon_{\mathbf k'} - \varepsilon_{\mathbf k}), 
\end{equation}
with 
\begin{equation}
   \widetilde{\mathcal W}_{\mathbf k',\mathbf k}
    = \frac{ 4\pi \bigl|\mathcal G_{\mathbf k',\mathbf k}\bigr|^2}{\sinh(\beta c_l |\mathbf{q}|)}.
    \label{Eq: StylishW}
\end{equation}
Finally,  the phase space integral reduces to an integration 
over the toroidal Fermi surface which, in the thin-torus limit $\kappa_\mathrm{F}/k_0 \ll 1$,
reads
\begin{equation}
    \int \frac{d^3\mathbf{k}'}{(2\pi)^3}  \delta( \varepsilon_{\mathbf k'} - \mu ) 
    \dots  \approx  \frac{\kappa_\mathrm{F} k_0}{(2\pi)^3 v_0} \int    d\varphi'd\theta'
    \dots,
    \label{phasespaceintegral}
\end{equation}
where we have used the Jacobian of the transformation to toroidal 
coordinates $\kappa'(k_0+\kappa' \cos \varphi')$, evaluated on the Fermi surface 
$\kappa'=\kappa_\mathrm{F}$, and 
the linear dispersion in Eq.~\eqref{lineardisp}; at leading order in $\kappa_\mathrm{F}/k_0$, 
we set  $k_0+ \kappa_\mathrm{F} \cos\varphi' \approx k_0$.
In this limit, the momentum transfer between states $\mathbf{k}$ and $\mathbf{k}'$
simplifies to (see Appendix~\ref{App: momentum transfer} for details)
\begin{equation}
|\mathbf{q}| = |\mathbf{k}-\mathbf{k}'| 
=
\sqrt{
4k_0^2\sin^2\!\left(\frac{\theta-\theta'}{2}\right)
+
4\kappa_\text{F}^2\sin^2\!\left(\frac{\varphi-\varphi'}{2}\right)
}.
\label{Eq: decay-rate5}
\end{equation}
Moreover, using Eq.~\eqref{Eq: overlap}, the Bloch spinor overlap $\langle u_{\mathbf{k}'} | u_\mathbf{k} \rangle $ 
depends only on the poloidal angle difference $\varphi-\varphi'$.
Therefore the scattering kernel $\widetilde{\mathcal W}_{\mathbf k',\mathbf k}$
depends on initial and final momenta only through toroidal and poloidal angle differences 
and is translationally invariant on the toroidal 
Fermi surface to leading order in $\kappa_\mathrm{F}/k_0$. 

Under the approximations discussed above, the linearized BTE can be solved exactly to leading 
order in $\kappa_\mathrm{F}/k_0$, as discussed next.

\subsection{Transport lifetimes}
\label{sec:tau_tr}

To solve Eq.~\eqref{Eq:BTE5} we make the ansatz
\begin{equation}
\varphi_{\mathbf{k}}
=
e  \mathbf{E} \cdot \mathbf{\Lambda}_\mathbf{k},
    \label{Eq:BTE6}
\end{equation}
where $\mathbf{\Lambda}_\mathbf{k}$ is the vector mean free path \cite{ZimanElectronPhonon}. 
Upon inserting Eq.~\eqref{Eq:BTE6} into
Eq.~\eqref{Eq:BTE5} and using Eqs.~\eqref{velocity}, ~\eqref{phasespaceintegral},
and the fact that the direction of $\mathbf E$ is arbitrary, we arrive at the integral equation 
on the Fermi torus
\begin{equation}
\hat e_\kappa = \frac{\kappa_\mathrm{F} k_0}{(2\pi)^3v_0^2} \int d\varphi' d\theta' \, 
\widetilde{\mathcal W}_{\mathbf k',\mathbf k} \left( \mathbf{\Lambda}_\mathbf{k}
-\mathbf{\Lambda}_{\mathbf{k}'} \right) .
    \label{Eq: BTE6bis}
\end{equation}
Since to leading order in $\kappa_\mathrm{F}/k_0$, 
the kernel $\widetilde{\mathcal{W}}_{\mathbf{k}',\mathbf{k}}$
depends only on angle differences, 
$\widetilde{\mathcal{W}}_{\mathbf{k}',\mathbf{k}}=
\widetilde{\mathcal{W}}(\theta-\theta',\varphi-\varphi')$, 
the collision integral is a convolution on the torus, so the collision operator is diagonal 
in the double Fourier basis $e^{im\theta+in\varphi}$. 
We show in Appendix~\ref{App: exact-solving-boltzmann} that the exact solution 
of Eq.~\eqref{Eq: BTE6bis} reads
\begin{equation}
    \Lambda^i_\mathbf{k} = v_0 \tau^{i}_\mathrm{tr} e^i_\kappa, \quad i=x,y,z,
    \label{Eq:solutionBTE}
\end{equation}
where the scattering lifetimes, evaluated at the chemical potential $\mu$,
are given by
\begin{align}
    \Gamma_{\rm{tr}}^{x,y} =& \frac{1}{\tau^{x,y}_{\mathrm{tr}}} =
 \frac{\kappa_\mathrm{F} k_0}{(2\pi)^3v_0}\int d\varphi' d\theta'  
     \,  \widetilde{\mathcal{W}}_{\mathbf k',\mathbf k}
      \left(1 - \cos \theta' \cos\varphi' \right),
    \label{Eq: transport-lifetime-integralx} \\
 \Gamma_{\rm{tr}}^{z} =&\frac{1}{\tau^z_{\mathrm{tr}}} =   
 \frac{\kappa_\mathrm{F} k_0}{(2\pi)^3v_0}\int d\varphi' d\theta'  
     \,  \widetilde{\mathcal{W}}_{\mathbf k',\mathbf k}
      \left(1 - \cos\varphi' \right).
  \label{Eq: transport-lifetime-integralz}
\end{align}
The angular factors in 
Eqs.~\eqref{Eq: transport-lifetime-integralx} 
and~\eqref{Eq: transport-lifetime-integralz} 
are Ziman transport factors~\cite{ZimanElectronPhonon}, 
adapted to the toroidal geometry: they suppress forward-scattering
processes that do not efficiently relax the current along a given direction and weight
more heavily momentum transfers that strongly deflect the velocity vector on the Fermi torus.
This directional dependence produces the anisotropy between in-plane $(x,y)$ and the axial $(z)$ transport lifetimes. 
The anisotropy  originates from the Fermi-surface geometry and is distinct from the 
material-dependent anisotropy of the energy dispersion close to the nodal line, 
which we neglect here (we set the dispersion anisotropy parameter $\alpha=1$). 

We use Eqs.~\eqref{Eq: transport-lifetime-integralx} and~\eqref{Eq: transport-lifetime-integralz} 
to evaluate the conductivity in Sec.~\ref{sec5}.

%%%%%%%%%%%%%

\section{Quasiparticle decay rate}
\label{sec4}

Before analyzing the conductivity, we evaluate the temperature dependence of the 
single-particle decay rate $\Gamma_\mathbf{k}(T)$, which can be accessed 
experimentally through the linewidth of ARPES peaks \cite{Mahan2000, Sobota2021, Bian2016, Hosen2017, Lv2021}. 
The decay rate follows from the out-scattering term of the BTE~\eqref{Eq:BTE5} 
and it is given by
\begin{equation}
    \Srate_\mathbf{k} = \int \frac{d^3\mathbf{k}'}{(2\pi)^3} \mathcal{W}_{\mathbf{k',k}} .
      \label{Eq: decay-rate1}
\end{equation}
\begin{table}[t]
    \centering
    \begin{tabular}{|c|c|c|} \hline 
         &   $\Srate $ & $\Gamma^i_{\rm{tr}}$\\ \hline 
 $T \ll \TBGpol $& $T^3$&$T^{5}$\\ \hline 
 $\TBGpol \ll T \ll \TBGtor$& $T^2$&$T^{2}$\\ \hline 
         $\TBGtor \ll T$& 
     $T$&$T$ \\ \hline\end{tabular}
    \caption{Temperature dependence of the decay rate $\Srate$ 
    and the scattering rates $\Gamma^{i}_{\rm{tr}}$ ($i=x,y,z$)
    for quasiparticles on the Fermi surface.  
    The corresponding resistivity behaves as $\rho\propto T^5$ for $T\ll \TBGpol$, $\rho\propto T^2$ 
    in the intermediate regime $\TBGpol \ll T \ll \TBGtor $, and $\rho \propto T$ at $T\gg \TBGtor$, 
    mirroring the standard BG behavior at low and high temperatures and the torus-induced intermediate scaling.}
    \label{tab:Table1}
\end{table}
Under the approximations discussed in Sec.~\ref{sec3}, $\Srate_\mathbf{k}$
can be expressed as the Fermi surface integral
\begin{align}
    \Srate = \frac{g_0 ^2 \kappa_\text{F} k_0  }{(2\pi)^2 \rho_M c_l v_0} \int \!
 d\varphi' d\theta'\frac{ 
    |\mathbf{q}| }{ \sinh (\beta c_l|\mathbf{q}| )}  
    \cos^2 \left(\frac{\varphi -\varphi'}{2}\right),
    \label{Eq: decay-rate4}
\end{align}
where the modulus of the momentum transfer $|\mathbf{q}|$ is given in Eq.~\eqref{Eq: decay-rate5}, 
and we have used the spinor overlap from Eq.~\eqref{Eq: wavefunction-in-toroidal-coordinates}. 
Since the integrand only depends on angle differences, $\Gamma_\mathbf{k}$ is uniform over the 
Fermi surface up to $\mathcal{O}(\kappa_\mathrm{F}/k_0)$ corrections, and we henceforth drop the $\mathbf{k}$ label.  
We then shift the integration variables 
and, using periodicity and reflection symmetry, restrict the integration domain.  
With the substitutions $\theta',\varphi'\rightarrow 2\theta'+\theta,2\varphi'+\varphi$, 
we arrive at the compact expression
\begin{align}
      \Srate = \frac{8g_0 ^2\kappa_{\mathrm{F}} k_0}{\pi^2\rho_M c_l v_0}  \int_0^{\frac{\pi}{2}}  
      d\varphi' \int_0^{\frac{\pi}{2}}d\theta' 
      \frac{Q(\theta',\varphi')  \cos^2 (\varphi')}{\sinh[2\beta c_l Q(\theta',\varphi')]}  , 
      \label{Eq: decay-rate6}
\end{align}
where we have defined 
\begin{equation}
Q(\theta,\varphi) = \sqrt{ k_0^2 \sin^2(\theta) + \kappa_{\mathrm{F}}^2 \sin^2 (\varphi)}. 
 \label{Q}   
\end{equation}

As discussed in Sec.~\ref{sec2:level4}, the thin-torus limit $\kappa_\text{F}\ll k_0$ implies
two BG temperature scales, $\TBGpol \propto \kappa_\text{F}$ and $\TBGtor \propto k_0$,
and therefore three temperature regimes. The evaluation of Eq.~\eqref{Eq: decay-rate6}
in these regimes is presented in Appendix~\ref{App: decay-rate};
here we summarize the resulting power laws in Table~\ref{tab:Table1} and provide simple phase-space 
arguments for the scalings.
\begin{figure}[!t]
    \centering
    \includegraphics[width=1\linewidth]{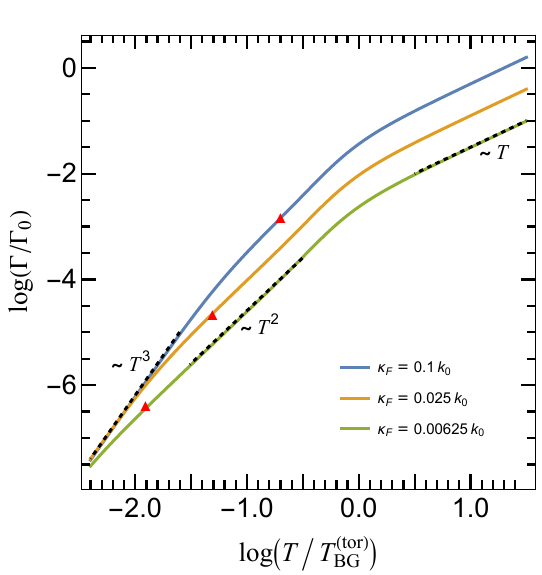}
    \caption{Temperature dependence of the quasiparticle decay rate $\Gamma(T)$ 
    obtained from a direct numerical evaluation of Eq.~\eqref{Eq: decay-rate4}.
    Results are shown for $c_l/v_0=10^{-2}$ and for several chemical
    potentials (see legend), with $\Gamma_0$ defined in Eq.~\eqref{Gamma0-defined}.
    The red markers indicate $T = \TBGpol$. 
    Dashed black lines represent asymptotic power-law fits across the three temperature regimes.}
    \label{fig: Numerics1}
\end{figure}
For $T\ll \TBGpol$, both angular separations are restricted by the thermal bound $Q\lesssim k_\mathrm{B} T/2c_l$, 
giving $\Delta\theta \sim k_\mathrm{B}T/(2c_l k_0)$ and $\Delta\varphi \sim k_\mathrm{B}T/(2c_l \kappa_\text{F})$. 
Hence the available phase space scales as $\Delta\theta \Delta \varphi \propto T^2$. 
Together with the explicit factor $Q\sim T$ in Eq.~\eqref{Eq: decay-rate6}, 
this gives $\Gamma\propto T^3$. 

In the intermediate window $\TBGpol \ll T\ll \TBGtor$,  
poloidal momentum transfers of order $\kappa_\text{F}$ are thermally allowed,
while the large toroidal radius $k_0$ forces the toroidal angular separation 
to be small in order to keep $Q\lesssim k_\mathrm{B}T/2c_l$:
$\Delta\theta \sim Q/k_0 \sim k_\mathrm{B} T/(2c_l k_0)\propto T$.
Equation~\eqref{Eq: decay-rate6} provides one explicit factor $Q\sim T$, 
and together with the restriction $\Delta\theta\propto T$, makes the available phase space on the torus $\Delta\theta \Delta \varphi \propto T$, yielding $\Gamma\propto T^2$.

For $T\gg \TBGtor$, the angular phase space saturates, and the leading temperature dependence 
arises from $1/\sinh x \simeq 1/x$ for small $x=2\beta c_l Q$ in Eq.~\eqref{Eq: decay-rate6}, 
yielding the scattering rate proportional to the number of phonons available for scattering, i.e., 
$\Gamma\propto T$.

More explicitly, we obtain the following asymptotic power laws for quasiparticles at the Fermi level  
($\varepsilon=\mu$), derived in Appendix~\ref{App: decay-rate}. In the low temperature regime, $T \ll \TBGpol$, 
the decay rate follows
\begin{equation}
    \Srate  =  \frac{7\zeta (3)}{4\pi}  \Gamma_0 \left( \frac{T}{\TBGtor} \right)^3 , 
    \label{decay-rate-lowT}
\end{equation}
where $\zeta (s)$ is the Riemann zeta function and
\begin{equation}
    \Gamma_0 = \frac{g_0^2 k_0^3}{\rho_M c_l v_0}.\label{Gamma0-defined}
\end{equation}
sets the characteristic electron-phonon scattering scale. In the intermediate regime  we find
\begin{equation}
    \Srate =  \frac{\pi}{16} \Gamma_0  \left( \frac{\TBGpol}{\TBGtor} \right) \left( \frac{T}{\TBGtor} \right)^2,
    \label{decay-rate-interT}
\end{equation}
and finally in the high-temperature regime
\begin{equation}
      \Srate =  \frac{1}{4}\Gamma_0   \left( \frac{\TBGpol}{\TBGtor} \right) \left( \frac{T}{\TBGtor} \right),
      \label{decay-rate-highT}
\end{equation}
the standard high-temperature linear-in-$T$ dependence~\cite{Ashcroft}. Figure~\ref{fig: Numerics1} shows a direct numerical evaluation of Eq.~\eqref{Eq: decay-rate4}, 
confirming the three scaling regimes and power laws summarized in Table~\ref{tab:Table1}.

%%%%%%%%%%%%%

\section{Conductivity}
\label{sec5}

Having established the temperature dependence of the quasiparticle decay rate $\Gamma(T)$, we now 
turn our attention to the dc conductivity, which is controlled by the transport lifetimes 
introduced in Sec.~\ref{sec3}.

The electric current density is
\begin{equation}
    \mathbf{J}
    = e \int \frac{d^3 \mathbf{k}}{(2\pi)^3}
        \mathbf v(\mathbf k) f_{\mathbf k}.
\end{equation}
Inserting Eq.~\eqref{Eq: linearizing-distribution-function} together with 
Eq.~\eqref{Eq:BTE6}, the equilibrium contribution vanishes by symmetry, yielding
\begin{equation}
    \mathbf J
    = -e^2 \int \frac{d^3 \mathbf{k}}{(2\pi)^3}
        \mathbf v(\mathbf k)
        \left( \mathbf E\cdot \mathbf{\Lambda}_\mathbf k \right)
        \frac{\partial n_{\mathrm F}(\varepsilon_{\mathbf k})}
             {\partial\varepsilon_{\mathbf k}}.
    \label{Eq: CurrentDensity-BTE-formula}
\end{equation}
Using the BTE solution \eqref{Eq:solutionBTE} and 
identifying $\mathbf{J} = \boldsymbol{\sigma}\,\mathbf{E}$, 
we obtain for the conductivity tensor
\begin{equation}
    \sigma_{ij} = \frac{e^2 \kappa_\mathrm{F} k_0 \tau^j_{\rm{tr}}}{(2\pi)^3}  \int  d\varphi  \, d\theta  
    \, e^i_\kappa e^j_\kappa ,
\end{equation}
which yields the longitudinal conductivities 
\begin{align}
     &\sigma_{xx} = \sigma_{yy} =  \frac{e^2 \tau^{x}_{\rm{tr}}\mu k_0}{8\pi }, \qquad
     \sigma_{zz}  = \frac{e^2 \tau^z_{\rm{tr}}\mu k_0}{4\pi },
     \label{dc-conductivity-formula-for-NLS}
\end{align}
while the off-diagonal terms vanish by symmetry. 
From this expression, we see that the temperature dependence of the conductivity 
is set by transport lifetimes $\tau^i_{\rm{tr}}$ given in Eqs.~\eqref{Eq: transport-lifetime-integralx} 
and~\eqref{Eq: transport-lifetime-integralz}. 

With manipulations similar to those leading to Eq.~\eqref{Eq: decay-rate6} we obtain
\begin{align}
      \Gamma_{\rm{tr}}^x &= \frac{8g_0 ^2 k_0\kappa_{\mathrm{F}} }{\pi^2\rho_M c_l v_0}  
      \int_0^{\frac{\pi}{2}}  d\varphi' \int_0^{\frac{\pi}{2}}d\theta' 
      \frac{Q(\theta',\varphi') \cos^2 (\varphi')}{\sinh\left[2\beta c_l Q(\theta',\varphi')\right]}  
      \times \nonumber \\
      &\times 
      \big[ 1-  \cos(2\varphi') \cos(2\theta ') \big], 
      \label{Eq: transport-lifetime-main-integral-x}
\end{align}
\begin{align}
      \Gamma_{\rm{tr}}^z &= \frac{8g_0 ^2 k_0\kappa_{\mathrm{F}} }{\pi^2\rho_M c_l v_0}  
      \int_0^{\frac{\pi}{2}}  d\varphi' \int_0^{\frac{\pi}{2}}d\theta' 
      \frac{Q(\theta',\varphi') \cos^2 (\varphi')}{\sinh\left[2\beta c_l Q(\theta',\varphi')\right]}  
      \times \nonumber \\
      &\times 
      \big[1- \cos(2\varphi') \big], 
      \label{Eq: transport-lifetime-main-integral-z}
\end{align}
with $Q(\theta,\varphi)$ defined in Eq.~\eqref{Q}. 
\begin{figure}[!t]
    \centering
    \includegraphics[width=1\linewidth]{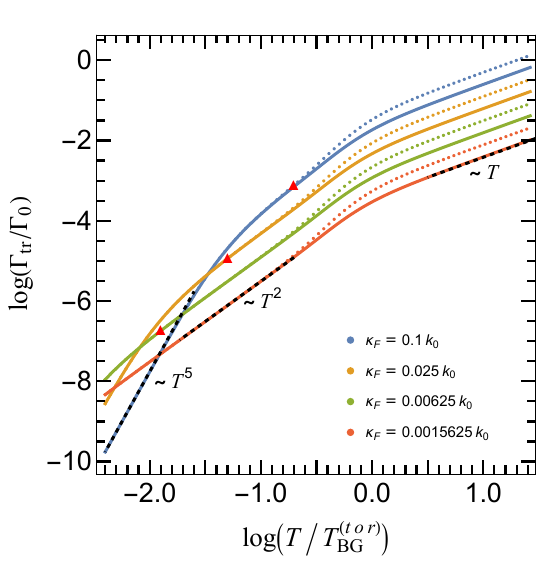}
    \caption{Temperature dependence of the transport scattering rates  $\Gamma_\text{tr}^z(T)$ (solid line) 
    and $\Gamma_\text{tr}^x(T)$ (dotted line) obtained from direct numerical evaluation of 
    Eqs.~\eqref{Eq: transport-lifetime-main-integral-x} and \eqref{Eq: transport-lifetime-main-integral-z}.
    Results are shown for $c_l/v_0=10^{-2}$ and for several chemical
    potentials (see legend), with $\Gamma_0$ defined in Eq.~\eqref{Gamma0-defined}.
    The red markers indicate $T = \TBGpol$. Note that for the red curve, the poloidal BG temperature $\TBGpol$ 
    falls outside the temperature window displayed. 
    Dashed black lines represent asymptotic power-law fits across the three temperature regimes.}
    \label{fig: Numerics2}
\end{figure}
In Appendix~\ref{App: Transport-lifetime-elastic} we analytically evaluate
Eqs.~\eqref{Eq: transport-lifetime-main-integral-x} and~\eqref{Eq: transport-lifetime-main-integral-z}, 
obtaining the temperature dependences summarized in Table~\ref{tab:Table1}. 
We see a clear agreement between the transport scattering rate results of  Table~\ref{tab:Table1} 
and the numerics in Fig.~\ref{fig: Numerics2}. 
We find that the temperature dependence of the transport scattering rate at 
low temperature follows
\begin{align}
       \Gamma_{\rm{tr}}^x =\Gamma_{\rm{tr}}^z = \frac{93\zeta(5)}{4\pi} \Gamma_0   \left( \frac{\TBGtor}{\TBGpol} \right)^2  
       \left( \frac{T}{\TBGtor} \right)^5, \label{TL-lowT}
\end{align}
where $\zeta (5) \simeq 1.037$. This $T^{5}$ temperature dependence of the rate 
implies  $\sigma_{ii}\propto T^{-5}$ for the conductivities, which is the standard 
low-temperature behavior in conventional metals \cite{ZimanElectronPhonon}. 
At intermediate temperatures, similar to Eq.~\eqref{decay-rate-interT}, 
we get
\begin{align}
       \Gamma_{\rm{tr}}^x =\Gamma_{\rm{tr}}^z= \frac{\pi  }{32} \Gamma_0  
       \left( \frac{\TBGpol}{\TBGtor} \right)   
       \left( \frac{T}{\TBGtor} \right)^2.
       \label{TL-interT}
\end{align}
%

%The angular transport weighting factors in Eqs.~\eqref{Eq: transport-lifetime-main-integral-x} 
%and~\eqref{Eq: transport-lifetime-main-integral-z} are 
%$\left(1 - \cos \theta' \cos\varphi' \right)$ and $\left(1 -  \cos\varphi' \right)$ respectively. 
In this regime, only $\Delta\theta  \propto T$ %(from the phase-space arguments above Eq.~\eqref{decay-rate-lowT}) 
while $\Delta \varphi$ spans the whole integration domain. 
As a result, the angular transport weighting factors in Eqs.~\eqref{Eq: transport-lifetime-main-integral-x} 
and~\eqref{Eq: transport-lifetime-main-integral-z} contribute at $\mathcal{O}(1)$,
yielding the same $T^2$ dependence for the transport scattering rate as for the decay rate.
Finally, in the high temperature regime $T \gg \TBGtor$, we find
\begin{align}
     \Gamma_{\rm{tr}}^z =\frac{\Gamma_{\rm{tr}}^x}{2} = \frac{1}{8} \Gamma_0  
      \left( \frac{\TBGpol}{\TBGtor} \right) 
       \left( \frac{T}{\TBGtor} \right),
       \label{TL-highT}
\end{align}
i.e., a linear-in-$T$ dependence. From Eq.~\eqref{dc-conductivity-formula-for-NLS}, we thus find that
\begin{align}
    &\frac{\sigma_{zz}}{\sigma_{xx}} = 2, & T \ll  \TBGtor, \\
    &\frac{\sigma_{zz}}{\sigma_{xx}} = 4, & T \gg  \TBGtor. 
\end{align}
The increase of the conductivity anisotropy upon raising the temperature through $\TBGtor$ is a direct consequence of 
the toroidal Fermi-surface geometry.

%%%%%%%%%%%%%

\section{Summary and Outlook}
\label{sec6}

In this work, we have analyzed quasiparticle scattering and charge transport in doped NLSs with a toroidal Fermi surface, 
focusing on longitudinal acoustic phonons coupled to electrons via a deformation potential. 
Using a low-energy two-band model with a circular nodal ring, 
we consider the regime below the Lifshitz scale, where the Fermi surface is a thin torus characterized 
by a major radius $k_0$ and a minor radius $\kappa_{\mathrm F}$. This geometry gives rise to two parametrically distinct 
BG temperature scales, $\TBGpol\propto c_l \kappa_{\mathrm F}$ 
and $\TBGtor \propto c_l k_0$, associated with phonons probing the poloidal and toroidal dimensions of the Fermi surface, respectively.
Using Fermi’s golden rule and the linearized BTE (which we solved exactly to leading order in $\kappa_\mathrm{F}/k_0$), 
we found that the quasiparticle decay rate and the conductivity exhibit three temperature regimes:
the low- and high-temperature limits reproduce the familiar BG behavior, while the intermediate $T^2/T^{-2}$
scaling emerges from the toroidal Fermi surface and the separation of the momentum scales $\kappa_{\mathrm F}$ and $k_0$.
Corrections of order $\kappa_\mathrm{F}/k_0$ break translational invariance on the torus and 
lead to additional angular harmonics in the collision kernel. However, we expect these subleading terms 
only renormalize the eigenvalues of the collision operator by $\mathcal{O}(\kappa_\mathrm{F}/k_0)$ 
and therefore are not expected to modify the power-law exponents of the leading terms 
that we obtain in the strict thin-torus limit. 

The origin of the modified scaling can be traced to the phase-space measure. For a spherical Fermi surface, the kinematic constraint $|\mathbf{q}| \lesssim T$ restricts both angular directions, giving a phase-space measure $\Delta \Omega \sim (\Delta \theta)^2 \sim T^2$. Combined with the typical momentum transfer $|\mathbf{q}| \sim T$ and the transport weighting factor $(1-\cos \theta) \sim T^2$, this yields the conventional low-temperature Bloch-Gr\"uneisen scaling $\propto T^5$. In contrast, for a toroidal Fermi surface in the thin-torus limit, the phase space measure factorizes as $\Delta\theta\, \Delta \varphi$, with $\Delta\theta \sim T$ while $\Delta\varphi$ spans the full integration domain in the intermediate regime. Together with $|\mathbf{q}| \sim T$ and transport weighting factors that contribute at $\mathcal{O}(1)$, this leads to the parametrically distinct $T^2$ scaling obtained here.

\begin{comment}
The origin of the modified scaling can be traced to the phase-space measure. For a spherical Fermi surface, the kinematic constraint $|\mathbf{q}| \lesssim T$ restricts both angular directions, giving a phase-space measure $\Delta \Omega \sim (\Delta \theta)^2 \sim T^2$. Combined with the typical momentum transfer $|\mathbf{q}|$ and the additional transport weighting factor $(1-\cos \theta) \sim T^2$. This yields the conventional low-temperature Bloch-Gr\"uneisen scaling $\propto T^5$. In contrast, for a toroidal Fermi surface in the thin-torus limit, the phase space separates as $\Delta\theta \Delta \varphi$, with $\Delta\theta \sim T$ while $\Delta\varphi$ can vary throughout the integration domain in the intermediate regime. Along with $|\mathbf{q}| \sim T$ and the transport weighting factors giving $\mathcal{O}(1)$ terms, leads to the parametrically distinct $T^2$ scaling of the transport lifetime obtained here.
\end{comment}

Our predictions can be tested experimentally. The quasiparticle decay rate is accessible 
in ARPES experiments as a linewidth, and in materials with a toroidal Fermi surface and linear dispersion 
along the minor-radius direction, we predict an intermediate window where $\Srate \propto T^2$. 
The corresponding temperature dependence of the transport lifetime, and thus the conductivity, can be probed via
standard four-probe measurements, for which we predict $\sigma \propto T^{-2}$, i.e., a resistivity $\rho \propto T^2$ 
in the same regime. Importantly, the onset of this intermediate regime is controlled by 
$\TBGpol \propto \kappa_\mathrm{F}\propto \mu$, so tuning the chemical potential should shift the 
lower crossover temperature, providing a direct way to distinguish this
phonon mechanism from other sources of $T^2$ resistivity. While a $T^2$ resistivity is commonly attributed to
intraband electron–electron scattering~\cite{Baber1937, Ashcroft}, our results show that in NLSs 
with a toroidal Fermi surface, a purely electron–phonon mechanism can generate the same temperature dependence. 

We also find that the Fermi-surface geometry leads to an enhanced conductivity anisotropy 
in the high-temperature regime.
We emphasize that, in the thin-torus limit, circular nodal-line semimetals exhibit 
an intrinsically anisotropic transport lifetime. Even when the scattering kernel depends 
only on angular differences, isotropic treatments are inadequate, and the transport lifetime 
must instead be evaluated using Eqs.~\eqref{Eq: transport-lifetime-integralx} 
and~\eqref{Eq: transport-lifetime-integralz}.

One of the key assumptions of our model is the absence of energy dispersion along the nodal line, 
i.e., the nodal line is assumed to be flat in energy. In general, no crystal symmetry enforces such flatness, 
and consequently most nodal-line semimetals do not exhibit simple toroidal Fermi surfaces. 
Instead, more complicated geometries such as Dupin cyclides~\cite{Ahn2017} arise, 
to which our analysis does not directly apply, since a clear separation between poloidal and toroidal 
radii is not possible and interband scattering between coexisting electron and hole pockets 
may become important. While materials such as $\mathrm{CaAgAs}$~\cite{Emmanouilidou2017} 
have been proposed to host nearly toroidal (donut-like) hole Fermi surface --- serving as a ``hydrogen atom'' 
realization of a nodal-line semimetal --- in many materials, such as $\mathrm{ZrSiS}$, 
achieving a purely toroidal Fermi surface imposes a lower bound on the achievable carrier 
density~\cite{Singha2017, Poulomi2024}.

In materials hosting additional trivial Fermi pockets or multiple nodal loops, 
these channels will contribute parallel conduction paths with their own, 
typically single-scale BG behavior. Nevertheless, as long as a toroidal pocket 
with $\kappa_\mathrm{F} \ll k_0$ remains well defined and dominates the low-energy density of states, 
the two-scale structure and the intermediate $T^2/T^{-2}$ regime should survive 
as robust features of the total response.

Several refinements and extensions are left for future work, such as including crystal anisotropy, 
energy-dispersive or non-circular nodal lines, additional Fermi pockets, 
and analysing how they renormalize prefactors 
and/or shift the crossover scales $\TBGpol$, $\TBGtor$. We expect the existence of two BG scales 
and an intermediate $T^2/T^{-2}$ window to be a generic consequence of Fermi surfaces with two widely 
separated characteristic momenta, even when the assumptions of an ideal circular nodal ring and 
a perfectly isotropic elastic medium considered here are relaxed. By contrast, quantitative features 
such as the precise values $\sigma_{zz}/\sigma_{xx} = 2$ at low $T$ and $4$ at high $T$ are 
geometry-specific and will be modified by crystal anisotropy and additional pockets.
It would also be interesting to extend our study to thermal transport and thermoelectric response, 
by evaluating the corresponding Boltzmann integrals for the heat current and extracting the temperature dependence of the thermal conductivity and the Seebeck coefficient across the two BG scales.
Including impurity and electron–electron scattering on an equal footing would 
enable a more quantitative comparison with experiments on specific materials.

%%%%%%%%%%%%%

\begin{acknowledgments}
We acknowledge helpful discussions with F. Buccheri, R. Citro, and R. Egger.
A.A. is grateful for the PhD studentship provided by the School of Science and Technology, 
City St George’s, University of London.
\end{acknowledgments}

\section*{Data availability}
The data underlying the figures in this paper are available in Zenodo \cite{Zenodo}.

%%%%%%%%%%%%%%%%%%%%%%%%%%%%%%%%%%%%%%%%%%%%%%%%%%%%%%%%%%%%%%%%%
% APPENDIX
%%%%%%%%%%%%%%%%%%%%%%%%%%%%%%%%%%%%%%%%%%%%%%%%%%%%%%%%%%%%%%%%%

\appendix
\setcounter{figure}{0}
\renewcommand{\thefigure}{A\arabic{figure}}

%%%%%%%%%%%%%

\section{Linearized Boltzmann equation}
\label{App:Boltzmann}

In this Appendix, we outline the derivation of the linearized BTE~\eqref{Eq:BTE5}.
Substituting the transition rates~\eqref{Eq:transition-rate-W} 
into Eq.~\eqref{Eq:BTE3}, we can write the integrand of the collision integral~\eqref{Eq:BTE3} as
\begin{widetext}
\begin{align}
        \mathcal{C}_{\mathbf{k},\mathbf{k}'} & \equiv W_{\mathbf{k},\mathbf{k}'} \, f_{\mathbf{k}'}(1-f_{\mathbf{k}}) -
        W_{\mathbf{k}',\mathbf{k}} \, f_{\mathbf{k}}(1-f_{\mathbf{k}'}), \nonumber \\
        &= 2\pi |{\mathcal{G}_{\mathbf{k},\mathbf{k}'}}|^2 \bigg\{ n_{\mathrm{B}}(\Omega_\mathbf{q} ) f_{\mathbf{k}'}(1-f_{\mathbf{k}}) 
        \, \delta (\varepsilon_\mathbf{k}  - \varepsilon_{\mathbf{k}'} - \Omega_\mathbf{q} )+ [n_{\mathrm{B}}(\Omega_\mathbf{q} ) + 1]
        f_{\mathbf{k}'}(1-f_{\mathbf{k}}) 
        \, \delta (\varepsilon_\mathbf{k}  - \varepsilon_{\mathbf{k}'} + \Omega_\mathbf{q} ) \nonumber \\
        &- n_{\mathrm{B}}(\Omega_\mathbf{q} ) f_{\mathbf{k}}(1-f_{\mathbf{k}'}) 
        \, \delta (\varepsilon_{\mathbf{k}'}  - \varepsilon_\mathbf{k} - \Omega_\mathbf{q} )- [n_{\mathrm{B}}(\Omega_\mathbf{q} ) + 1]
        f_{\mathbf{k}} (1-f_{\mathbf{k}'}) 
        \, \delta (\varepsilon_{\mathbf{k}'}  - \varepsilon_\mathbf{k} + \Omega_\mathbf{q} ) \bigg\} ,
\label{Eq: AppA1}
\end{align}
where we have used the symmetry $\left|{\mathcal{G}_{\mathbf{k},\mathbf{k}'}}\right| = \left|{\mathcal{G}_{\mathbf{k}',\mathbf{k}}}\right|$. 
Grouping terms with the same delta-function argument and linearizing the distribution function using 
Eq.~\eqref{Eq: linearizing-distribution-function}, we obtain
\begin{align}
      \mathcal{C}_{\mathbf{k},\mathbf{k}'} &=2\pi |{\mathcal{G}_{\mathbf{k},\mathbf{k}'}}|^2 
      \Bigg\{ 
      \left[ 
      \big( n_{\mathrm{B}}(\Omega_\mathbf{q} )+1-n_{\mathrm{F}} (\varepsilon_{\mathbf{k}'})\big)
      \varphi_{\mathbf{k}} \frac{\partial n_{\mathrm{F}} (\varepsilon_\mathbf{k})}{\partial \varepsilon_\mathbf{k}}
      -
      \big( n_{\mathrm{B}}(\Omega_\mathbf{q} )+n_{\mathrm{F}} (\varepsilon_\mathbf{k}) \big)
      \varphi_{\mathbf{k}'} 
      \frac{\partial n_{\mathrm{F}} (\varepsilon_{\mathbf{k}'})}{\partial \varepsilon_{\mathbf{k}'}} \right]
      \delta(\varepsilon_{\mathbf{k}'}  - \varepsilon_\mathbf{k} + \Omega_\mathbf{q}) \nonumber\\
    & + \left[ \big( n_{\mathrm{B}}(\Omega_\mathbf{q} )+ n_{\mathrm{F}} (\varepsilon_{\mathbf{k}'}) \big) 
    \varphi_{\mathbf{k}} 
    \frac{\partial n_{\mathrm{F}} (\varepsilon_\mathbf{k})}{\partial \varepsilon_\mathbf{k}} 
    - \big( n_{\mathrm{B}}(\Omega_\mathbf{q} )+1-n_{\mathrm{F}} (\varepsilon_\mathbf{k})\big)
    \varphi_{\mathbf{k}'} 
    \frac{\partial n_{\mathrm{F}} (\varepsilon_{\mathbf{k}'})}{\partial \varepsilon_{\mathbf{k}'}} \right] 
    \delta(\varepsilon_{\mathbf{k}'}  
    - \varepsilon_\mathbf{k} - \Omega_\mathbf{q})\Bigg\},
        \label{Eq: AppA2}
\end{align}
where the zeroth-order terms cancel by detailed balance.
Next, we use the two identities 
\begin{align}
 \left[n_{\mathrm{B}}(\Omega_\mathbf{q} )+n_{\mathrm{F}} (\varepsilon_\mathbf{k}) \right] 
     \frac{\partial n_{\mathrm{F}} (\varepsilon_{\mathbf{k}'})}{\partial \varepsilon_{\mathbf{k}'}} =   
     \left[ n_{\mathrm{B}}(\Omega_\mathbf{q} ) + 1- n_{\mathrm{F}}(\varepsilon_{\mathbf{k}'} ) \right] 
  \frac{ \partial n_{\mathrm{F}} (\varepsilon_\mathbf{k})}{\partial \varepsilon_\mathbf{k}} 
    &, &\mathrm{when}\; \varepsilon_{\mathbf{k}'}  = \varepsilon_\mathbf{k} - \Omega_\mathbf{q} , \label{Eq: AppA3}\\
 \left[ n_{\mathrm{B}}(\Omega_\mathbf{q} )+1-n_{\mathrm{F}} (\varepsilon_\mathbf{k})\right] 
 \frac{\partial n_{\mathrm{F}} (\varepsilon_{\mathbf{k}'})}{\partial \varepsilon_{\mathbf{k}'}}=
   \left[  n_{\mathrm{B}}(\Omega_\mathbf{q} ) + n_{\mathrm{F}}(\varepsilon_{\mathbf{k}'}  ) \right]
   \frac{\partial n_{\mathrm{F}} (\varepsilon_\mathbf{k})}{\partial \varepsilon_\mathbf{k}}
    &, &\mathrm{when}\; \varepsilon_{\mathbf{k}'}  = \varepsilon_\mathbf{k} + \Omega_\mathbf{q} ,
     \label{Eq: AppA4}
\end{align}
and recast Eq.~\eqref{Eq: AppA2} into the form
\begin{align}
      \mathcal{C}_{\mathbf{k},\mathbf{k}'} &=2\pi |{\mathcal{G}_{\mathbf{k},\mathbf{k}'}}|^2 
      \frac{\partial n_{\mathrm{F}} (\varepsilon_\mathbf{k})}{\partial \varepsilon_\mathbf{k}}
      \Bigg\{ 
      \big[ n_{\mathrm{B}}(\Omega_\mathbf{q} ) + 1 - n_{\mathrm{F}}(\varepsilon_{\mathbf{k}'}) \big]  
      \delta(\varepsilon_{\mathbf{k}'}  - \varepsilon_\mathbf{k} + \Omega_\mathbf{q}) 
     +  \big[ n_{\mathrm{B}}(\Omega_\mathbf{q}) + n_{\mathrm{F}}(\varepsilon_{\mathbf{k}'})\big]
        \delta (\varepsilon_{\mathbf{k}'}  - \varepsilon_\mathbf{k} - \Omega_\mathbf{q}) \Bigg\}  
        \left( \varphi_{\mathbf{k}} - \varphi_{\mathbf{k}'}  \right) \nonumber \\
        & =  
      \frac{\partial n_{\mathrm{F}} (\varepsilon_\mathbf{k})}{\partial \varepsilon_\mathbf{k}}
      \mathcal{W}_{\mathbf{k}',\mathbf{k}}  \left( \varphi_{\mathbf{k}} - \varphi_{\mathbf{k}'}  \right),
    \label{Eq: AppA5}
\end{align}
with the kernel $\mathcal{W}_{\mathbf{k}',\mathbf{k}} $ given in Eq.~\eqref{Eq: StylishW1}.
\end{widetext}
We can now write the BTE~\eqref{Eq:BTE2} to first order in the deviation from equilibrium as
\begin{align}
    e\mathbf{E} \cdot \partial_{\mathbf{k}} n_{\mathrm{F}} (\varepsilon_\mathbf{k})= 
    \int \frac{d^3\mathbf{k}'}{(2\pi)^3}\mathcal{C}_{\mathbf{k},\mathbf{k}'}.
    \label{Eq: AppA6}
\end{align}
Substituting Eq.~\eqref{Eq: AppA5} in Eq.~\eqref{Eq: AppA6}, we finally get
\begin{equation}
     e\mathbf{E} \cdot \mathbf{v}(\mathbf{k}) = \mathcal{J}_{\mathbf{k}},
\end{equation}
where $\mathcal{J}_{\mathbf{k}}$ is defined in Eq.~\eqref{Eq:BTE4}.

%%%%%%%%%%%%%%

\section{Solution of the BTE}
\label{App: exact-solving-boltzmann}

In this Appendix, we present the solution of the linearized BTE~\eqref{Eq: BTE6bis}.
The equation can be cast in the form
\begin{equation}
    e_\kappa^i(\theta,\varphi) 
    = \int d\theta' d\varphi' 
    \, K(\theta,\varphi;\theta',\varphi') \left[ \Lambda^i (\theta,\varphi)-\Lambda^i (\theta',\varphi') \right],
    \label{Eq: exact-boltz-6}
\end{equation}
where the kernel $K(\theta,\varphi;\theta',\varphi')$ is given by
\begin{equation}
    K(\theta,\varphi;\theta',\varphi') = \frac{\kappa_\mathrm{F}k_0}{(2\pi)^3 v_0^2} 
    \, \widetilde{\mathcal{W}}_{\mathbf k',\mathbf k},
\end{equation}
and the angles $(\theta,\varphi)$ and $(\theta',\varphi')$ parametrize the momenta $\mathbf{k}$ and $\mathbf{k}'$
on the Fermi surface. As discussed in the main text, in the thin-torus limit, the kernel depends only on angle differences: 
\begin{equation}
     K (\theta,\varphi;\theta',\varphi') = \f(\theta-\theta',\varphi-\varphi'),
\end{equation}
where $f(\theta,\varphi)$ is a real function even in both arguments.
As a consequence, the integral operator in Eq.~\eqref{Eq: exact-boltz-6} 
takes the form of a convolution, and the equation can be solved by Fourier transform.

Expanding $e^i_\kappa$, $\Lambda^i$, and $f$ in a double Fourier series as
\begin{align}
      e^i_\kappa (\theta,\varphi) & = \sum_{m,n \in \mathbb{Z}} e^i_{\kappa,mn} e^{im\theta +in\varphi},
    \label{Eq: exact-boltz-7} \\ 
    \Lambda^i (\theta,\varphi) & = \sum_{m,n \in \mathbb{Z}}\Lambda^i _{mn} e^{im\theta +in\varphi}, \\ 
    \f(\theta,\varphi) & = \sum_{m,n \in \mathbb{Z}} \f_{mn} e^{im\theta +in\varphi},
\end{align}
and substituting the Fourier expansions into Eq.~\eqref{Eq: exact-boltz-6} yields
\begin{equation}
    e^i_{\kappa,mn}  =  (2\pi)^2 \left( \f_{00}- \f_{mn} \right) \Lambda^i _{mn} ,
\end{equation}
and thus
\begin{equation}
 \Lambda^i_{mn} =    \frac{e^i_{\kappa,mn}}{(2\pi)^2 \left( \f_{00}- \f_{mn} \right)}.
\end{equation}
Since $e^i_\kappa$ are simple trigonometric functions (see Eq.~\eqref{Eq: toroidal-unit-vector-definition}),
only a few Fourier coefficients are non-vanishing. 
Then, using  the even parity of $f$:
\begin{equation}
    \f_{m,n} = \f_{-m,n} = \f_{m,-n} = \f_{-m,-n} ,
\end{equation}
we arrive at
\begin{align}
    \Lambda^{i}(\theta,\varphi) &= \frac{ e_{\kappa}^i(\theta,\varphi)}{(2\pi)^2\left(\f_{00}- \f_{11} \right)}, \qquad i=x,y\\
    \Lambda^z (\theta,\varphi) & = \frac{ e_{\kappa}^z(\theta,\varphi)}{(2\pi)^2\left( \f_{00}- \f_{01}\right)}.
\end{align}
where
\begin{equation}
\f_{mn} = \int \frac{d\varphi \, d\theta}{(2\pi)^2} \f(\theta,\varphi)  \cos (m\theta) \cos(n\varphi). 
\end{equation}
This yields the transport lifetime expressions in 
Eqs.~\eqref{Eq: transport-lifetime-integralx} 
and \eqref{Eq: transport-lifetime-integralz}.

%%%%%%%%%%%%%%%
\section{Momentum transfer}
\label{App: momentum transfer}

In this Appendix, we derive the thin-torus expression for the magnitude of the momentum transfer 
between two states on the toroidal Fermi surface, Eq.~\eqref{Eq: decay-rate5}. 
Using the parameterization in Eqs.\eqref{Eq: torus-para-kx}--\eqref{Eq: torus-para-kz}, 
the momentum transfer $\mathbf{q} = \mathbf{k}'-\mathbf{k}$ can be decomposed as
\begin{equation}
    \mathbf{q} = \mathbf{q}_0 + \mathbf{q'}
    \label{Eq: DR4}
\end{equation}
where the major-radius contribution is 
\begin{align}
    \mathbf{q}_0 & = k_0\left(\cos \theta' - \cos \theta \right)\hat{x}+  k_0
    \left( \sin \theta' - \sin \theta\right)\hat{y}, \label{Eq: DR5}
\end{align}    
and the minor-radius correction reads
\begin{align}
     \mathbf{q'} & = \kappa_{\mathrm{F}}\left( \cos \varphi' \cos \theta' - \cos \varphi 
    \cos \theta\right)\hat{x}+\kappa_{\mathrm{F}}( \cos \varphi' \sin \theta' -\nonumber \\
    &- \cos \varphi \sin \theta )\hat{y} + \kappa_{\mathrm{F}}
    \left(  \sin \varphi'-\sin \varphi\right) \hat{z}.
    \label{Eq: DR6}
\end{align}
A straightforward algebra then gives
\begin{align}
    |\mathbf{q}|^2 = 4k_0^2 \sin^2 \left(\frac{\theta -\theta'}{2} \right) 
    \bigg[ 1+\frac{1}{2} \frac{\kappa_{\mathrm{F}}}{k_0} (\cos \varphi' + \cos \varphi) +\nonumber \\
   +\left( \frac{\kappa_{\mathrm{F}}}{k_0} \right)^2\cos \varphi' \cos \varphi\bigg] +
   4\kappa_{\mathrm{F}}^2 \sin^2 \left(\frac{\varphi -\varphi'}{2} \right).\label{Eq: DR7}
\end{align}
To leading order in  $\kappa_{\mathrm{F}}/k_0\ll 1$ we may drop the terms in square brackets beyond unity,
which leads to Eq.~\eqref{Eq: decay-rate5}. 

%%%%%%%%%%%%%%%
\section{Decay rate}
\label{App: decay-rate}

In this Appendix, we present details of the analytical 
evaluation of the quasiparticle decay rate discussed in Sec.~\ref{sec4}.
We use Eq.~\eqref{Eq: decay-rate6} to extract asymptotic forms of $\Gamma(T)$ 
in the relevant temperature regimes.

%%%%%%%
\subsection{Low temperature}
\label{App: decay-rate-a}

In the regime $T \ll \TBGpol$, i.e., $\beta c_l k_0\gg  \beta c_l \kappa_{\mathrm{F}} \gg 1$, 
the factor $1/\sinh(2\beta c_l Q)$ in Eq.~\eqref{Eq: decay-rate6} exponentially 
suppresses contributions from all but the smallest momentum transfers.
Hence, the dominant contribution arises from the vicinity $\theta',\varphi'\simeq 0$, 
where we may linearize the sine terms by using $\sin x\simeq x$ and set $\cos \varphi'\simeq 1$. 
This yields
\begin{equation}
    \Srate = B  
    \int_0^{\frac{\pi}{2}}  d\varphi' \int_0^{\frac{\pi}{2}}d\theta' 
    \frac{\sqrt{ k_0^2 \theta'^2 + \kappa_{\mathrm{F}}^2 \varphi'^2}}{\sinh \left(2\beta c_l
    \sqrt{k_0^2 \theta'^2 + \kappa_{\mathrm{F}}^2 \varphi'^2} \right)}, 
    \label{Eq: DR-a1}
\end{equation}
with 
\begin{equation}
B= \frac{8g_0^2 \kappa_\mathrm{F}k_0}{\pi^2 \rho_Mc_lv_0}.
\label{Eq: DR3}
\end{equation}
Introducing the dimensionless variables 
\begin{equation}
   x = 2\beta c_l k_0 \theta', \qquad y = 2\beta c_l \kappa_{\mathrm{F}} \varphi',
\label{changeofvariables} 
\end{equation}

we obtain
\begin{equation}
    \Srate = \frac{B}{(2\beta c_l)^3\kappa_\mathrm{F}k_0}
    \int_0^{\beta c_l \kappa_{\mathrm{F}}\pi} \!\!\!\! dy \int_0^{\beta c_l k_0\pi} 
    \!\!\!\! dx \frac{\sqrt{x^2 +y^2}}{\sinh\sqrt{x^2 +y^2}}.  
    \label{Eq: DR-a2}
\end{equation}
Since $\beta c_l k_0, \beta c_l \kappa_{\mathrm{F}} \gg 1$, the upper integration limits 
can be extended to $\infty$. 
Switching to polar coordinates then gives
\begin{equation}
    \Srate =  \frac{B}{(2\beta c_l)^3\kappa_\mathrm{F}k_0} 
    \frac{\pi}{2}\int_0^{\infty} \; dr 
    \frac{r^2}{\sinh r}.
    \label{Eq: DR-a3}
\end{equation}
Using the standard integral $\int_{0}^{\infty} dr\, r^2/\sinh r = \frac{7}{2}\zeta(3)$, 
we finally obtain Eq.~\eqref{decay-rate-lowT}.
 
%%%%%%%

\subsection{Intermediate temperature}
\label{App: decay-rate-b}

In the intermediate regime $\TBGpol \ll T \ll \TBGtor$, i.e., 
$\beta c_l k_0\gg 1 \gg \beta c_l \kappa_{\mathrm{F}} $,
we may still linearize $\sin \theta'$ but not $\sin \varphi'$. 
Using Eq.~\eqref{changeofvariables} in Eq.~\eqref{Eq: decay-rate6} gives
\begin{align}
    \Srate = \frac{B}{(2\beta c_l)^2k_0}  \int_0^{\frac{\pi}{2}}  
    d\varphi' \int_0^{\beta c_lk_0 \pi } \!\!\!\! dx \frac{\cos^2 (\varphi')
    \sqrt{ x^2+b^2(\varphi')}}{\sinh\sqrt{ x^2 +b^2(\varphi')}} , 
    \label{Eq: DR-b2}
\end{align}
where $b(\varphi') = 2\beta c_l \kappa_{\mathrm{F}} \sin(\varphi')$. 
To lowest order in $\kappa_\mathrm{F}/k_0$ we may set $b(\varphi')\simeq 0$,
and since $\beta c_lk_0\gg 1$ we extend the upper limit of the $x$ integral to $\infty$.
This gives
\begin{align}
    \Srate = \frac{B}{(2\beta c_l)^2k_0}
    \int_0^{\frac{\pi}{2}}  d\varphi' \cos^2 (\varphi')
    \int_0^{\infty} dx \frac{x}{\sinh x} .
    \label{Eq: DR-b3}
\end{align}
Using $\int_{0}^{\infty} dx\, x/\sinh x = \frac{\pi^2}{4}$ and $\int_0^{\pi/2} d\varphi' \cos^2\varphi'=\pi/4$, 
we finally obtain Eq.~\eqref{decay-rate-interT}.

%%%%%%%
\subsection{High temperature}
\label{App: decay-rate-c}
Finally, in the regime $\TBGpol  \ll \TBGtor \ll T$, i.e., 
$1 \gg\beta c_l k_0\gg \beta c_l \kappa_{\mathrm{F}} $, 
we may linearize the denominator in Eq.~\eqref{Eq: decay-rate6} obtaining
\begin{equation}
        \Srate = B  \int_0^{\frac{\pi}{2}}  d\varphi' \cos^2 (\varphi')\int_0^{\frac{\pi}{2}}d\theta' 
        \frac{1}{2\beta c_l},
        \label{Eq: DR-c2}
\end{equation}
which immediately gives Eq.~\eqref{decay-rate-highT}.

%%%%%%%%%%%%%%

\section{Transport lifetimes}
\label{App: Transport-lifetime-elastic}

In this Appendix, we evaluate Eqs.~\eqref{Eq: transport-lifetime-main-integral-x} 
and \eqref{Eq: transport-lifetime-main-integral-z} in the three temperature regimes, 
following the same procedure as in Appendix~\ref{App: decay-rate}.

\subsection{Low temperature}
\label{App: transport-lifetime-a}

When $T \ll \TBGpol$, we expand the trigonometric functions to lowest order, obtaining ($i=x,y,z$)
\begin{align}
   \Gamma_{\rm{tr}}^i = 2B \int_{0}^{\frac{\pi}{2}}  \!\!\!\! d\varphi'
   \int_0^{\frac{\pi}{2}} \!\!\!\! d\theta' 
   \frac{(\varphi'^2 + a_i\, \theta'^2) \sqrt{ k_0^2 \theta'^2 + \kappa_{\mathrm{F}}^2 \varphi'^2}}{\mathrm{sinh}
   \left(2\beta c_l\sqrt{ k_0^2 \theta'^2 + \kappa_{\mathrm{F}}^2 \varphi'^2}\right)}, 
   \label{Eq: trans-life-a1-x}
\end{align}
where $B$ is defined in Eq.~\eqref{Eq: DR3} and $a_{x,y}=1, a_z=0$.
After the change of variables~\eqref{changeofvariables}, the integral becomes
\begin{align}
   \Gamma_{\rm{tr}}^i &= \frac{2B }{(2\beta c_l)^5\kappa^3_\mathrm{F}k_0}  
   \int_{0}^{\beta c_l \kappa_{\mathrm{F}} \pi}  \!\!\!\! dy \int_0^{\beta c_l k_0 \pi} 
   \!\!\!\! dx \frac{ \sqrt{x^2+y^2}}{\mathrm{sinh}\sqrt{x^2 + y^2}}  
   \nonumber \\
   & \times \left( y^2 + a_i \left( \frac{\kappa_{\mathrm{F}}}{ k_0}\right)^2 x^2   \right). 
    \label{Eq: trans-life-a2-x}
\end{align}
Extending the integration limits to $\infty$ and using
\begin{align}
     \int_0 ^{\infty} dy \int_0 ^{\infty} dx \frac{x^2 \sqrt{x^2 + y^2}}{\mathrm{sinh}\sqrt{x^2 + y^2}} 
     = \frac{93\pi \zeta(5)}{8}, 
     \label{Eq: trans-life-a4}
\end{align}
we obtain
\begin{align}
\Gamma_{\rm{tr}}^i = \frac{93 \pi\zeta(5)B }{4(2\beta c_l)^5\kappa^3_\mathrm{F}k_0}  
   \left[ 1 + a_i \left( \frac{\kappa_{\mathrm{F}}}{ k_0}\right)^2  \right].
   \label{Eq: trans-life-a5}
\end{align}
Neglecting the term of order $(\kappa_\mathrm{F}/k_0)^2$ gives Eq.~\eqref{TL-lowT}.\\

\subsection{Intermediate temperature}
\label{App: transport-lifetime-b}

When $\TBGpol \ll T \ll \TBGtor$, we may still linearize $\sin \theta'$, 
but not $\sin \varphi'$. Using Eq.~\eqref{changeofvariables} gives
\begin{align}
   \Gamma_{\rm{tr}}^i =  \frac{2B}{(2\beta c_l)^2k_0} 
   \int_{0}^{\frac{\pi}{2}}  d\varphi' \int_0^{\beta c_l k_0 \pi}\!\!\!\! dx
   \frac{\sqrt{ x^2 +b^2(\varphi')}}{\mathrm{sinh}\sqrt{ x^2+b^2(\varphi')}}   \nonumber \\
    \times\cos^2(\varphi')\left[ \sin^2(\varphi') + a_i \left(\frac{x}{2\beta c_l k_0} \right)^2
    \cos (2\varphi')   \right], 
    \label{Eq: trans-life-b1-x}
\end{align}
with $b(\varphi') = 2\beta c_l \kappa_{\mathrm{F}} \sin (\varphi')$. 
As done in Eq.~\eqref{Eq: DR-b2}, we neglect $b(\varphi')$ here as well and extend the limit of $x$-integration, obtaining
\begin{align}
   \Gamma_{\rm{tr}}^i &= \frac{2B}{(2\beta c_l)^2k_0} 
   \int_{0}^{\frac{\pi}{2}}  d\varphi' \int_0^{\infty}dx\frac{x}{\mathrm{sinh} \;x}  \cos^2(\varphi')  \nonumber \\
  & \times \left[ \sin^2 (\varphi') + a_i \left(\frac{x}{2\beta c_l k_0} \right)^2 \cos (2\varphi')  \right]. 
   \label{Eq: trans-life-b2-x}
\end{align}
Performing the $\varphi'$ and $x$ integrations yields
\begin{align}
  \Gamma_{\rm{tr}}^i = \frac{\pi^3 B}{32 (2\beta c_l)^2k_0} 
   \left[ 1 + a_i \left(\frac{\pi}{2\beta c_l k_0} \right)^2   \right].
   \label{Eq: trans-life-b3}
\end{align}
In this regime, we see that the second term in $\Gamma_{\rm{tr}}^i$ is suppressed 
by the factor $(\beta c_l k_0)^{-2} \ll 1$ and can be safely neglected, yielding Eq.~\eqref{TL-interT}.

%%%%%%%

\subsection{High temperature}
\label{App: transport-lifetime-c}

When $\TBGpol  \ll \TBGtor \ll T$, we linearize the denominator by using $\sinh x \approx x$ and obtain
\begin{align}
   \Gamma_{\rm{tr}}^i &=  2B 
   \int_{0}^{\frac{\pi}{2}}  d\varphi' \int_0^{\frac{\pi}{2}} d\theta' \frac{1}{2\beta c_l}  
   \cos^2(\varphi')  \nonumber \\
   &\times \left[ \sin^2 \varphi' +  a_i \cos(2\varphi')\sin^2\theta' \right] . 
   \label{Eq: trans-life-c1-x}
\end{align}
Performing the $\varphi'$ and $\theta'$ integrations gives Eq.~\eqref{TL-highT}.\\

%%%%%%%%%%%%%%

\bibliography{bib_file}

\end{document}